\documentclass[twocolumn]{article}
\usepackage{amssymb}
\usepackage{abstract}
\usepackage[margin=71pt]{geometry}
\usepackage{cite}
\usepackage{graphicx}
\usepackage{color}

\newcommand{\ignore}[1]{}
\newtheorem{problem}{Open problem}
\newtheorem{defn}{Definition}

\let\oldbibliography\thebibliography
\renewcommand{\thebibliography}[1]{%
  \oldbibliography{#1}%
  \setlength{\itemsep}{0pt}%
}

\title{\vspace{-20pt}Diffusion in the Lorentz gas}
\author{Carl P. Dettmann\\\em School of Mathematics, University of Bristol, Bristol BS8 1TW, UK}
\date{\today}

\begin{document}
\twocolumn[
  \begin{@twocolumnfalse}
    \maketitle
%\begin{abstract}
The Lorentz gas, a point particle making mirror-like reflections
from an extended collection of scatterers, has been a useful model of
deterministic diffusion and related statistical properties for over a century.
This survey summarises recent results, including periodic and aperiodic models, finite
and infinite horizon, external fields, smooth or polygonal obstacles,
and in the Boltzmann-Grad limit.  New results are given for several moving particles
and for obstacles with flat points.  Finally, a variety of applications are presented.
\vspace*{20pt}
%\end{abstract}
  \end{@twocolumnfalse}
]
\setcounter{tocdepth}{2}
\tableofcontents

\section{Introduction}
The Lorentz gas was proposed by H. A. Lorentz in 1905~\cite{Lorentz05}
to model thermal and electrical conductivity of metals.  The interactions between
electrons were neglected, and the ions considered fixed, so the model consists
of a single moving particle in an extended
array of fixed scatterers.  A Boltzmann-like approximation of uncorrelated
collisions was made, so the implicit assumption was that the scatterers were
of low density.  Physically this makes sense (though
ignoring electron-electron interactions and quantum effects) if the scatterers
are reinterpreted as lattice defects rather than ions.  

The subsequent century, especially the last decade, has seen a wealth of
results on related models, with periodic, quasiperiodic and aperiodic scatterer
arrangements, two, three and more dimensions, internal and external forces,
and many other generalisations.  Lorentz models have illuminated relevant
fields, both mathematical (probability and dynamical systems) and in the
physical sciences (foundations of statistical mechanics, molecular simulation,
scattering and transport in periodic and random environments).

This review gives an overview of the latest developments,
in particular since a previous survey by the same author~\cite{D00}; see also
Refs.~\cite{CD06,Klages07,Chernov10}. There are also new calculations in
Sec.~\ref{s:nondis} and (mostly old) open problems highlighted
throughout.  The order is logical rather than historical, starting with the widely
investigated and relatively well understood periodic models and moving towards
previously studied random models.  The final section draws together some
relevant applications.  Corresponding quantum/wave systems are a huge
and omitted field; see for example Refs.~\cite{Stockmann07,VH09,JJWM11}.

\section{Preliminaries}
\subsection{Microscopic dynamics}\label{s:micro}
\begin{figure}
\centerline{\includegraphics[width=300pt]{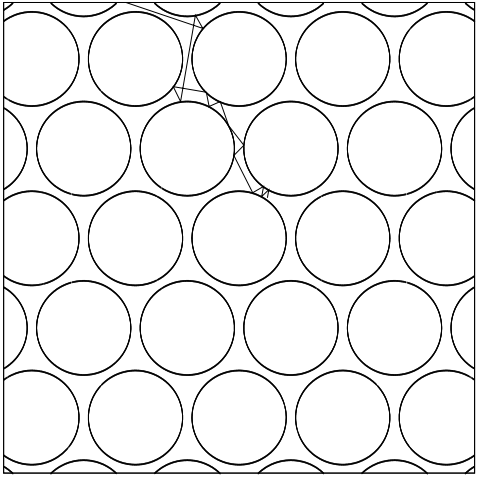}}
\vspace{-25pt}
\caption{A triangular periodic Lorentz gas. This has finite horizon (see Sec.~\protect\ref{s:hor}),
as there are no collision-free trajectories.  Infinite horizon may be obtained with smaller scatterers
or a square lattice.\label{f:lor}}
\end{figure}

A point particle with location ${\bf x}(t)\in\mathbb{R}^d$
as a function of time $t$, moves freely except for reflections from an
infinite collection of scatterers $D_i\subset\mathbb{R}^d$;  See Fig.~\ref{f:lor}.
Free motion is at constant velocity of unit magnitude (without loss of generality), so
\begin{equation}
\frac{d{\bf x}}{dt}={\bf v},\qquad |{\bf v}|=1.
\end{equation}
This equation is solved together with that for the boundary to determine the
time of next collision, a quadratic equation for the most common case of
scatterers which are balls. After reaching a scatterer, a reflection
takes place according to the usual rule that angle of reflection equal
angle of incidence
\begin{equation}\label{e:coll}
{\bf v}_+={\bf v}_--2\frac{{\bf v}_-\cdot{\bf n}}{{\bf n}\cdot{\bf n}}{\bf n}
\end{equation}
where a centred dot denotes the usual scalar product and $\bf n$ is a
vector normal to the boundary.  This formula does not require that
$\bf n$ be a unit vector, hence avoiding calculation of a square root.

Apart from the unbounded domain, the dynamics is exactly that of a
mathematical billiard~\cite{CM06,Tabachnikov05,Gutkin12}.  Basic properties of
billiards which are useful in the Lorentz gas context are the existence of a
uniform equilibrium measure, proportional to Lebesgue measure in position
($\mathbb{R}^d$) and velocity ($\mathbb{S}^{d-1}$) spaces, and invariant
under the dynamics in the continuous time dynamics (billiard flow).  For the
billiard map (dynamics from one collision to the next) the corresponding measure is
uniform on the boundary of the scatterer(s) and on the projection of velocity
parallel to the boundary.  With respect to this equilibrium measure there is
an exact formula for the mean flight time between collisions
\begin{equation}
\langle\tau\rangle=\frac{|Q|S_{d-1}}{|\partial Q|V_{d-1}}=\frac{|Q|}{|\partial Q|}
\frac{2\sqrt{\pi}\Gamma(\frac{d+1}{2})}{\Gamma(\frac{d}{2})}
\end{equation}
where $|Q|$ and $|\partial Q|$ are the volume and surface of the billiard respectively, while $V_{d}$
and $S_{d-1}$ are the volume and surface of the $d$-dimensional unit ball respectively. Angle
brackets denote expectation.  For $d=2$ we find $\pi|Q|/|\partial Q|$ and for $d=3$ we find
$4|Q|/|\partial Q|$.

Another property derived from billiards is the symplectic structure of the dynamics, leading to a symmetric
Lyapunov spectrum in the usual Hamiltonian fields (no field, electric and/or magnetic fields) and with Gaussian or
Nos\'e-Hoover thermostats~\cite{MD98,WL98}.  Also, the dynamics is time-reversible~\cite{RQ92}, in that 
in all these cases except the magnetic field there is an involution $\iota$ reversing the dynamics:
$\iota\circ \Phi^t\circ \iota=\Phi^{-t}$. For the flow $\iota$ is defined by reversal of velocity,
$\iota({\bf x},{\bf v})=({\bf x},-{\bf v})$ , and for the map by the reflection law,
$\iota({\bf x},{\bf v}_-)=({\bf x},{\bf v}_+)$ from Eq.~(\ref{e:coll}).  Here, $\Phi^t$ denotes the evolution
forward by time $t\in\mathbb{R}$ for the flow or collisions $t\in\mathbb{Z}$ for the map.

Dynamical properties of billiards depend on the geometry of the boundary. We will
assume the following unless stated explicitly:

\begin{defn}
Dispersing billiard: All scatterers are disjoint, convex with strictly
positive curvature and $C^3$ smooth.
\end{defn}
 
These requirements ensure that a parallel beam of initial conditions
spreads out at each collision, leading to strongly chaotic properties
(Sec.~\ref{s:dyn}).  Various generalisations of the dynamics and dispersing
condition are considered in later sections.

\subsection{Diffusion}\label{s:diff}
The main macroscopic property considered has been diffusion.  For heat
conduction see Ref.~\cite{LWWZ05} and references therein, but note that
the Lorentz gas collisions do not transfer energy, and so local thermal
equilibrium is not generally satisfied~\cite{DD99}.  Viscosity has also
been considered~\cite{BS96}.

The displacement $\Delta(t)={\bf x}(t)-{\bf x}(0)$ is a deterministic
function of the initial position and velocity.  Considering a probability
measure on the set of initial conditions and/or scatterer configurations,
we can study the distribution of $\Delta(t)$ including its moments as a
function of $t$.  Using
$i,j\in\{1,\ldots d\}$ as spatial indices, we can seek the
following limiting properties, in roughly increasing order of strength:
\begin{description}
\item[Current]
\begin{equation}
{\bf J}=\lim_{t\to\infty}\frac{1}{t}\langle\Delta\rangle
\end{equation}
If there is no external field, then the current is clearly zero due to
time reversibility.  If there is an external field, the following
properties should be defined in terms of $\Delta-{\bf J}t$.
\item[Mean square displacement]
\begin{equation}\label{e:msd}
D_{ij}=\lim_{t\to\infty}\frac{1}{2t}\langle\Delta_i\Delta_j\rangle
\end{equation}
where $D_{ij}$ is the diffusion matrix/tensor.
\item[Central limit theorem]
\begin{equation}
\frac{\Delta(t)}{\sqrt{t}}\Rightarrow {\cal N}(0,2D_{ij})
\end{equation}
where $\cal N$ is the multivariate normal distribution.
\item[Brownian motion]
\begin{equation}
\frac{\Delta(st)}{\sqrt{t}}\Rightarrow W(s)
\end{equation}
for $s\in[0,1]$, where $W$ is the standard $d$-dimensional Wiener process
with covariance matrix $2D_{ij}$.
\item[Local limit theorem]
Given an unbounded sequence of times $t_n$ and scatterers with displacements
$\Delta_n/t_n\to {\bf x}$ and Voronoi cells (excluding the scatterers
themselves) of volume $V_n$, the probability $P_n$ of reaching the cell at
time $t_n$ has the expected limit:
\begin{equation}
\frac{t_n^{d/2}P_n}{V_n}\to \phi({\bf x})
\end{equation}
where $\phi$ is the density function of ${\cal N}(0,2D_{ij})$.
\end{description}
Here, $\Rightarrow$ denotes convergence in distribution as $t\to\infty$.
The factors of two are required so that $D_{ij}$ is the coefficient
in the corresponding hydrodynamic equation
\begin{equation}\label{e:hydro}
\frac{\partial}{\partial t}\rho=D_{ij}\frac{\partial}{\partial x_i}
\frac{\partial}{\partial x_j}\rho
\end{equation}
where $\rho({\bf x},t)$ is the density of particles.  In cases with
sufficient symmetry (eg triangular periodic or isotropic random Lorentz
gases), the diffusion matrix is $D_{ij}=D\delta_{ij}$ with $D$ the
diffusion coefficient. For more details see Sec.~2 of Ref.~\cite{D12}.
Note that anomalous versions of the above results hold in some
situations; see Sec.~\ref{s:DIH} for details.

\subsection{Burnett coefficients}\label{s:Burnett}
It is sometimes useful to consider higher order diffusion processes.
Generalising Ref.~\cite{vB82} slightly to allow for the non-isotropic case,
we Fourier transform the density and expand in a formal power series,
using subscripts for spatial indices.  The derivation is given in two dimensions
for clarity, but easily generalises to arbitrary dimension.
\begin{eqnarray}
&&F(k_1,k_2,t)\label{e:Qdef}\\
&=&\int\int d\Delta_1 d\Delta_2 e^{-ik_1\Delta_1-ik_2\Delta_2}\rho(\Delta_1,\Delta_2,t)\nonumber\\
&=&\sum_{n_1=0}^{\infty}\sum_{n_2=0}^\infty \frac{(-i)^{n_1+n_2}}{n_1!n_2!}k_1^{n_1}k_2^{n_2}
\langle\Delta_1^{n_1}\Delta_2^{n_2}\rangle\nonumber\\
&=&\exp\sum_{n_1=0}^{\infty}\sum_{n_2=0}^\infty \frac{(-i)^{n_1+n_2}}{n_1!n_2!}k_1^{n_1}k_2^{n_2}
\langle\Delta_1^{n_1}\Delta_2^{n_2}\rangle_c\nonumber
\end{eqnarray}
by expanding the exponential in the first equation.  The final equality defines the cumulants
$\langle\rangle_c$, which have the important property that they are additive for independent random
variables.  We will use the notation $M_{ab}=\langle\Delta_1^a\Delta_2^b\rangle$ for
the moments and $Q_{ab}=\langle\Delta_1^a\Delta_2^b\rangle_c$ for the cumulants.  For example
if odd moments are zero due to symmetry we have
\begin{eqnarray}\label{e:Q}
Q_{00}=0,\qquad\qquad\;\;\;&&Q_{20}=M_{20},\\
Q_{40}=M_{40}-3M_{20}^2,&&Q_{22}=M_{22}-M_{20}M_{02}.\nonumber
\end{eqnarray}

Differentiating Eq.~(\ref{e:Qdef}), we find
\begin{equation}
F_t=F\sum_{n_1=0}^{\infty}\sum_{n_2=0}^\infty \frac{(-i)^{n_1+n_2}}{n_1!n_2!}k_1^{n_1}k_2^{n_2}\partial_tQ_{n_1n_2}
\end{equation}
where the late time limit (if it exists) of the derivative is given by
\begin{equation}
\lim_{t\to\infty}\partial_tQ_{n_1n_2}=\lim_{t\to\infty}\frac{1}{t}Q_{n_1n_2}
\end{equation}
Thus we have
\begin{equation}
\lim_{t\to\infty}\frac{F_t}{F}=-\sum_{mn}D_{mn}k_mk_n+\sum_{mnpq}B_{mnpq}k_mk_nk_pk_q+\ldots
\end{equation}
with $B$ the tensor form of the Burnett coefficient, as in Refs.~\cite{Gaspard05,D03}.
Thus the Burnett coefficient can be interpreted as a fourth derivative
term in Eq.~(\ref{e:hydro}).

Note that there are exactly the same number of independent $B$ and $Q$ coefficients
for any dimension and level of symmetry; the even coefficients are related by
\begin{eqnarray}
D_{11}&=&\lim_{t\to\infty}\frac{1}{2t}Q_{20},\qquad\quad\;\\
B_{1111}&=&\lim_{t\to\infty}\frac{1}{24t}Q_{40},\qquad
B_{1122}=\lim_{t\to\infty}\frac{1}{24t}Q_{22}.\nonumber
\end{eqnarray}
This is a generalisation of Eq.~(\ref{e:msd}); for the limit
to exist, the term $3M_{20}^2$ in Eq.~(\ref{e:Q}) of order $t^2$
needs to cancel an equivalent contribution from $M_{40}$ to give a
result of order $t$.  Thus Burnett coefficients can be anomalous even when
diffusion is normal.

The cumulants beyond second order are exactly zero for a normal distribution, and so
quantify approach to it.  The Burnett coefficients appear explicitly in corrections of
local limit theorems in the case of independent random variables~\cite{Petrov75}.

\subsection{Periodicity and horizons}\label{s:hor}
Periodic scatterer configurations are natural from both mathematical
and physical perspectives.  Mathematically, a periodic Lorentz gas is
a $\mathbb{Z}^d$ cover over a billiard in a torus, from which many
useful properties may be derived; in particular the equilibrium measure
on the torus excluding the scatterer(s) is finite.

Physically, the Lorentz gas with circular/spherical scatterers of radius
$R$ is obtained by considering molecular dynamics of two particles of
radius $R/2$ with periodic boundary conditions in relative coordinates
(that is, removing the uniform centre of mass motion)~\cite{D00}.

There is, however, an important issue to consider with periodic
models.  For the simplest case of a square lattice with non-overlapping
circular scatterers, the particle can move freely parallel to a unit
lattice vector without ever colliding with a scatterer; this property
is called  {\bf infinite horizon}.  For a triangular lattice with
sufficiently large scatterers (as in Fig.~\ref{f:lor})
there is no such trajectory and the 
time between collisions is bounded; this is {\bf finite horizon}.
In aperiodic models (Sec.~\ref{s:aper} below)
a third possibility exists, where there is neither
an infinite trajectory nor a bound on the collision time; this is
{\bf locally finite horizon}.

In three dimensions there may be cylinders and/or slabs of trajectories
with no collisions, corresponding to ``cylindrical'' or ``planar''
infinite horizons; see Refs.~\cite{D12,NSV12,Sanders08}.  In general $d\geq 2$
we consider the lattice ${\cal L}$ of translations of the Lorentz gas.
Following Refs.~\cite{D12,NSV12} a free subspace is an inextensible linear
subspace $V\subset\mathbb{R}^d$ such that for a point
${\bf x}\in\mathbb{R}^d$ the set ${\bf x}+V$ does not intersect any
scatterer (but may be tangent); $V$ is a lattice subspace.  The
corresponding horizon is constructed by obtaining the maximal connected
set $B_H\subset{\bf x}+V^\perp$ containing $\bf x$ that has $V$ as a
free subspace, where $V^\perp$ is the linear space perpendicular to $V$.
The horizon itself is the set
\begin{equation}
H=\{({\bf x},{\bf v}):{\bf x}\in B_H+V,{\bf v}\in V\cap\mathbb{S}^{d-1}\}
\end{equation}
where the sphere $\mathbb{S}^{d-1}$ imposes the restriction to unit speed.
The dimension of the horizon $d_H$ is the dimension of $V$, and satisfies
$1\leq d_H\leq d-1$.  A {\bf maximal horizon} is a horizon of maximal dimension
for a given Lorentz gas, a {\bf principal horizon} is of dimension $d-1$, and
an {\bf incipient horizon} is one in which $B_H$ is zero $d-d_H$ dimensional
measure, for example a plane tangent to scatterers on both sides.

Let the free flight function $F(t)$ be the probability (given initial conditions chosen
according to the equilibrium measure) of not colliding before time $t$.
Also, let $F_H(t)$ be the probability of remaining in the spatial projection
of the horizon $B_H+V$ for time $t$.  Then~\cite{D12}
\begin{equation}\label{e:F(t)}
F_H(t)=
\frac{S_{d_H-1}\int_{B_H}\int_{B_H}
\Delta^{\rm vis}({\bf x},{\bf y})d{\bf x}d{\bf y}}
{S_{d-1}{\cal V}_H^\perp(1-{\cal P})t^{d-d_H}}
\end{equation}
where $\Delta^{\rm vis}$ counts the number
of ways $\bf x$ and $\bf y$ can be connected by a straight line entirely in
$B_H$. ${\cal V}_H^\perp$ is the volume of $V^\perp/{\cal L}_V^\perp$ where
${\cal L}_V^\perp$ is lattice obtained by the projection of ${\cal L}$
on $V^\perp$.  Finally ${\cal P}$ is probability that an arbitrary point
lies inside a scatterer.

Conjectured in Ref.~\cite{D12} and proven in the preprint~\cite{NSV12}:
\begin{equation}\label{e:F}
F(t)\sim\sum_{H\in\mathbb{H}}F_H(t)\qquad t\to\infty
\end{equation}
where $\mathbb{H}$ is the set of maximal horizons if at least one is
non-incipient.  In the limit $r\to 0$ the number of horizons diverges;
a non-rigorous calculation using Mellin transforms yields for a $d$-dimensional
cubic lattice~\cite{D12}
\begin{equation}\label{e:Fsmall}
\lim_{t\to\infty}tF(t)=\frac{\pi^{\frac{d-1}{2}}}{2^dd\Gamma(\frac{d+3}{2})\zeta(d)r^{d-1}}
+O(r^{\frac{1}{2}-\delta})
\end{equation}
for $\delta>0$ subject to the Riemann Hypothesis, the major unsolved problem
in number theory that asserts that the Riemann zeta function $\zeta(s)$ has no
zeros with real part greater than $1/2$~\cite{Conrey03}.

Also conjectured in Ref.~\cite{D12} (except for the
explicit exponent in the third case) and proven in~\cite{NSV12}:
\begin{equation}
F(t)\asymp\left\{\begin{array}{cc}
t^{-2}&3\leq d\leq 5\\
t^{-2}\ln t&d=6\\
t^{\frac{2+d}{2-d}}&d>6
\end{array}\right.
\end{equation}
if there is at least one incipient (but no actual) principal horizon.  The explicit exponent
matches the numerical fits in Ref.~\cite{D12} for $d\leq 8$ beyond which the latter are
not reliable.  Note that the numerical simulations were carried out for cubic Lorentz
gases with scatterers just touching, thus violating the dispersing condition.

\begin{problem}
What is the form of $F(t)$ if the maximal horizon is incipient but
not principal?
\end{problem}

In three or more dimensions, no periodic arrangement of spheres with a
single sphere per unit cell has finite horizon.
It is possible to create finite horizon configurations by having
non-spherical scatterers, for example by considering a generic 
lattice and shrinking the Voronoi tessellation slightly to create
strictly convex scatterers, or by a sufficiently large number of
randomly placed spheres per unit cell.

\begin{problem}
Find an explicit periodic arrangement of equal sized
non-overlapping spheres with finite horizon in dimension $d\geq 3$.
\end{problem}

\section{Periodic with finite horizon}
\subsection{Dynamical properties}\label{s:dyn}
Dynamical properties of motion in a two dimensional torus with strictly
convex obstacles have been known since Sinai's 1970 paper~\cite{Sinai70}
which showed that such systems exhibit the Kolmogorov property, which
implies both ergodic properties (for example ergodicity and mixing) and
hyperbolic properties (for example positive Lyapunov exponents for almost
every initial condition).  A modern detailed treatment of these questions
for more general billiards may be found in Ref.~\cite{CM06}.

Hyperbolicity holds for higher dimensional dispersing billiards (hence
Lorentz gases), though in this case the structure of the singularities
(due to tangential orbits) is much more involved~\cite{BCST03,BT12}.
These results extend automatically to the extended (Lorentz gas) case.

Ergodicity has been shown where the scatterers are algebraic varieties
(such as spheres)~\cite{BCST02}, or when the growth of singularities is
less than exponential (and hence dominated by the exponential stretching
associated with hyperbolicity)~\cite{BT08,BBT08} however this condition
known not to be always satisfied~\cite{BT12}.  These difficulties were
not appreciated before the turn of the millennium and so earlier results
on higher dimensional Lorentz gases need to be treated with caution.
The following assertion is however very likely true:

\begin{problem}
Are all dispersing billiards on $\mathbb{T}^d$ with $d\geq 3$ and finite
horizon ergodic?
\end{problem}

Ref.~\cite{BT08,BBT08} proves a stronger ergodic property, that of exponential
mixing of the billiard map, under this condition (sub-exponential complexity).
This had been shown earlier for two dimensional billiards by Young~\cite{Young98}.
the rate of mixing for the flow is more difficult; the best results are
stretched exponential in two dimensions~\cite{Chernov07}, with exponential
only for billiards with a non-eclipsing condition (hence a finite number
of scatterers in $\mathbb{R}^2)$~\cite{Stoyanov01}; progress has been made
on non-billiard models that are hyperbolic with singularities~\cite{BL12}.

A stronger result in another direction is the Bernoulli property for the
map and flow, shown for all billiards in arbitrary dimension with
non-zero Lyapunov exponents for both map and flow, and that satisfy the
K-property~\cite{CH96}.

\subsection{Transport}
Now we consider the dynamics in the extended space.  In $d\leq 2$, the random
walk is well known to be recurrent, with trajectories returning infinitely
often arbitrarily close to their starting point.  This holds also for the
two dimensional Lorentz gas, with the stronger property of ergodicity in
the full space shown in Refs.~\cite{Conze99,Schmidt98}.  Clearly this is
not expected in higher dimensions.

The two dimensional Lorentz gas has been shown to satisfy all the diffusive
properties given in Sec.~\ref{s:diff}: Bunimovich and Sinai showed convergence
to Brownian motion in 1981~\cite{BS81}, the local limit theorem was
proved by Sz\'asz and Varju in 2004~\cite{SV04} and the vector-valued
almost sure invariance principle by Melbourne and Nicol in 2009~\cite{MN09}.
In higher dimensions, the central limit theorem was shown in Ref.~\cite{BT08}
under the condition of subexponential complexity. 

\begin{problem}
Are convergence to a Wiener process and local limit theorems satisfied for
$d\geq 3$?
\end{problem}

Noting $\Delta_i=\int_0^t v_i(s)ds$ we arrive at the expression~\cite{Bleher92}
\begin{equation}\label{e:GK}
\frac{1}{2}\langle\Delta_i\Delta_j\rangle
=t\int_0^t\langle v_i(0)v_j(s) \rangle ds
-\int_0^ts\langle v_i(0)v_j(s) \rangle ds
\end{equation}
which after division by $t$ and taking the limit gives the continuous
time version of the Green-Kubo formula for the diffusion coefficient $D$;
similar expressions apply to other transport coefficients such as viscosity
and heat conductivity~\cite{EM08}.  The equivalent expression for the
fourth order Burnett coefficients involves four-time correlation
functions~\cite{Gaspard05}.

Convergence of the integral follows from sufficiently rapid decay of the
velocity autocorrelation function; this was shown to be (at least) a
stretched exponential in Ref.~\cite{BS81}; similar multiple correlations
were used in Refs.~\cite{CD00,D03} to show that Burnett coefficients of
all orders also exist in the two dimensional case.

There is no known closed form expression for the diffusion coefficient $D$,
so there have been several analytical and numerical studies to approximate it.
In the limit of small gaps between scatterers the time between moving from
one gap to another is typically long and hence uncorrelated~\cite{MZ83}.
More sophisticated models extend this to take into account some
correlations, for example Markov chains with finite
memory~\cite{KD00,GS09,AM12}.  There is also a study of three dimensional
Lorentz gases using this approach~\cite{GNS11} including both finite and
infinite horizon regimes (Sec.~\ref{s:IH} below).

Alternative expressions for $D$ exist in terms of periodic orbits of the torus
dynamics (which are either periodic or translating in the full space).
Considering only orbits up to a maximum length yields an approximation to
the diffusion coefficient; see Refs.~\cite{CGS92,MR94,CEG95}.  A rigorous
basis of periodic orbit expansions is provided in Ref.~\cite{Pollicott91}.  

There are simpler models of deterministic diffusion, for example one
dimensional piecewise linear maps. Some models have a dense set of
parameter values at which there is a finite Markov partition and hence
the diffusion coefficient may be determined exactly using Markov chains
or periodic orbits~\cite{KD95,KD99,KK11,KGDK12}.  The diffusion coefficient
is a fractal function of parameters (denoted $\lambda$), for which the
non-smoothness has been fairly well quantified~\cite{KHK08}, being slightly
less smooth than Lipshitz.  An upper bound of the variation is
\begin{equation}
|D(\lambda)-D(\lambda')|<K|\lambda-\lambda'|(1+\ln|\lambda-\lambda'|)^2
\end{equation}
while a lower bound includes a single power of the logarithm. The
density dependence of the diffusion coefficient in the Lorentz gas is
expected to be smoother (since the flow is continuous if the pre- and
post- collisional states are considered connected); discussion and numerical
results were presented in Ref.~\cite{KD00} and for a ``flower'' Lorentz
gas in Ref.~\cite{HKG02}.
\begin{problem}
How smooth is the diffusion coefficient of the Lorentz gas as a function
of density?
\end{problem}

\subsection{External fields}\label{s:ext}
\subsubsection{Weak field and thermostat}\label{s:gauss}
The diffusion can also be calculated as the zero field limit of the
non-equilibrium conductivity~\cite{MH87} (Eq.~(\ref{e:JD}) below),
a standard approach for transport
coefficients in molecular simulation~\cite{EM08}.  In this case an
electric field is imposed, that provides a constant force on the particle
(assumed charged, though still not interacting with other moving particles).
In order to prevent an unbounded growth of the particle's energy, a
thermostat force is also applied, as commonly used in molecular
simulation~\cite{MD98,EM08}.  The most commonly used thermostat in this
context is the Gaussian isokinetic thermostat, for which the equation of
motion is
\begin{equation}
\frac{d{\bf v}}{dt}={\bf E}-\frac{{\bf E}\cdot{\bf v}}{{\bf v}\cdot{\bf v}}
\end{equation}
where $\bf E$ is the constant electric field; the mass and charge are assumed
equal to unity.  This equation has the following properties~\cite{MD98}:
\begin{itemize}
\item The kinetic energy ${\bf v}\cdot{\bf v}/2$ is conserved (hence the
designation ``isokinetic'') and so it is usually assumed that the velocity
has unit magnitude.
\item The involution $\iota({\bf x},{\bf v})=({\bf x},-{\bf v})$ reverses the motion
(see Sec.~\ref{s:micro}).
\item The dynamics is conformally symplectic~\cite{DM96b,WL98}.  This means
that in two dimensions there is a conformal transformation to a field-free
billiard~\cite{WL98} and that in higher dimensions there is a symmetry of
the Lyapunov spectrum, sometimes called the conjugate pairing rule~\cite{DM96a,WL98}.
Before the latter was shown, the three dimensional 
Lorentz gas was used as a convenient system for numerical tests~\cite{DM95}.
\end{itemize}

If the electric field is to the right (${\bf E}=E{\bf e}_x$) and the direction
of motion ${\bf v}=(\cos\theta,\sin\theta)$ in the $(x,y)$ plane (without loss
of generality), the equation of motion reduces to $d\theta/dt=-F\sin\theta$
with solution
\begin{eqnarray}
\tan\frac{\theta}{2}&=&\tan\frac{\theta_0}{2}\exp\left[-\frac{t-t_0}{E}\right]\\
x&=&x_0-\frac{1}{E}\ln\frac{\sin\theta}{\sin\theta_0}\\
y&=&y_0-\frac{\theta-\theta_0}{E}
\end{eqnarray}
Note that the motion in $y$ (ie transverse to the field) is bounded by $\pi/E$.
The equation determining the collision with a spherical scatterer is
transcendental, however the shortest distance to a scatterer is a rigorous
lower bound on the time, and leads to a quadratically convergent numerical
algorithm~\cite{DM95}.

The two-dimensional finite horizon non-equilibrium Lorentz gas was considered
in Ref.~\cite{CELS93}, where it was shown that for sufficiently small field
the system remains ergodic, with a measure that is supported on the full phase
space but singular (multifractal) with respect to the equilibrium measure.
They give rigorous proofs of relations long stated in non-equilibrium physics,
for example the existence of a well defined current ${\bf J(E)}$ (the average
velocity) related in the limit of small field to the diffusion coefficient
(or in low symmetry cases, tensor) by the Einstein relation
\begin{equation}\label{e:JD}
{\bf J}=D{\bf E}+o({\bf E})
\end{equation}
Also, the diffusion tensor is continuous at zero field, the sum of the
Lyapunov exponents comes to
\begin{equation}
\lambda_++\lambda_-=-{\bf J}\cdot{\bf E}
\end{equation}
and the information dimension of the flow (Hausdorff dimension of the measure)
is given by the application of the Kaplan-Yorke-Young formula~\cite{Young82},
\begin{equation}\label{e:DI}
D_1=2+\frac{\lambda_+}{|\lambda_-|}=3-\frac{{\bf J}\cdot{\bf E}}{\lambda_0}+
o({\bf E}^2)
\end{equation}
where $\lambda_0$ is the magnitude of either of the Lyapunov exponents at zero field.
Recently, Ref.~\cite{BCKL12a} shows that the spatial projection of the ergodic measure
is absolutely continuous.  These authors also used a Gaussian thermostat as a coupling
mechanism in a multiparticle driven Lorentz gas~\cite{BCKL12b}.

\subsubsection{Other weak fields and thermostats}\label{s:weak}
Ref.~\cite{CELS93} incorporates a more general case of a weak magnetic field;
since the magnetic force is perpendicular to the velocity it does not affect the strength of
the thermostat and hence the sum of the Lyapunov exponents.  It does however break
the time reversibility (see Sec.~\ref{s:micro}). These results have been generalised in a
number of papers: Ref.~\cite{Chernov01} considers more general small forces,
Refs.~\cite{CD10,Zhang11} considers perturbed reflection functions, and Ref.~\cite{CZZ13}
considers more general models with small forces and collision perturbations satisfying
a conserved energy with compact phase space (cf the Galton board below) and
time reversibility.  Ref.~\cite{CZZ13} extends the perturbations to include shifted, rotated
or deformed scatterers and/or shifts at collision.  The results are similar, including
generalisations of the Einstein formulas.  There are also studies of electric and magnetic
fields in random Lorentz gases, discussed below in Sec.~\ref{s:random}.

From the point of view of molecular dynamics, there are a number of other thermostats
in use~\cite{MD98,JR10}.  The Nos\'e-Hoover thermostat retains reversibility and symplectic properties,
while adding a degree of freedom.  Its application to the Lorentz gas was considered numerically
in Ref.~\cite{RKH00}, leading again to multifractal attractors but somewhat different bifurcation
structure at strong field.  It was used as an example of fluctuation theorems in
Refs.~\cite{DK05,Gilbert06}.

Other deterministic thermostatting approaches have been less common, lacking time-reversibility
and Hamiltonian structure.  The constant friction thermostat was considered in Ref.~\cite{HM92} as
an example of a non-reversible model, again studying the multifractal attractors.  Note that there
are some subtleties to the definition of reversibility; it turns out that if the attracting fixed point
is excluded, the constant friction harmonic oscillator dynamics $(\dot{x},\dot{v})=(v,-x-\alpha v)$
may be reversed by the involution
\begin{equation}
\iota(x,v)=\frac{(x,-v-\alpha x)}{x^2+\frac{(v+\alpha x/2)^2}{1-\alpha^2/4}}
\end{equation}

A reversible thermostat-like model was considered numerically in Ref.~\cite{BR01}.  There, the motion
was as in a billiard except that when the particle leaves via a boundary, returning on the opposite side,
a dilation was applied, giving an effective field.  A steady current was observed, linear in the field
strength for weak field.  Fluctuation relations studied.  Infinite horizon and polygonal variants were
also considered (see later sections).

\subsubsection{Strong field and thermostat}
Some statements can be made about the Gaussian thermostat for strong
fields.  In the two-dimensional case, the conformal transformation to a
field-free billiard~\cite{WL98} guarantees ergodicity as long as the
appropriate conditions are satisfied: $|{\bf E}|<\kappa_{\rm min}$ where
$\kappa_{\rm min}$ is the minimum curvature of the scatterers ensures
that the dispersing condition is met; the finite horizon condition on the
transformed billiard also needs to be checked.   Smoothness of the current
as a function of field is also of long interest:
\begin{problem}
How smooth is the current as a function of field?
\end{problem}
A plot of this function is given in Ref.~\cite{D00}. For larger fields,
ergodicity appears to be broken by one of two mechanisms, elliptic stability
of a periodic orbit (breaking of the above condition) or crisis where an
attractor and its time reverse at high fields merge as the field is 
reduced~\cite{DM96c,OM99}.  At high fields the attractor(s) may be fractal or
stable periodic orbits.

\subsubsection{The Galton board}
The Galton board or quincunx, predates the Lorentz gas~\cite{Galton},
consisting of a periodic Lorentz gas with a constant
field but no thermostat.  It was proposed and is still used as a mechanical
demonstration of the binomial distribution, in which the particle falls under
the action of a gravitational field through a triangular lattice of
obstacles, moving left or right with approximately equal and independent
probabilities (due to the rapid decay of correlations).  The mechanical models
have various sources of friction, but it is clear that the idealised model
has unbounded kinetic energy as the particle continues to fall.

This model was studied rigorously in Ref.~\cite{CD09a} with some unexpected
results: The position of the particle grows on average as $t^{2/3}$, so
there is no linear drift.  From conservation of energy this means the speed
grows as $t^{1/3}$ (these had been obtained previously in the physics
literature~\cite{PL79,KR97}).  However the motion is also recurrent: The particle
returns infinitely many times to the vicinity of its starting point.  Note
that the region containing the slowest motion is not a small perturbation
of the field-free Lorentz gas, and so needs to be excluded (for example it
is possible to have an elliptic periodic orbit bouncing between two
scatterers).

\section{Periodic with infinite horizon}\label{s:IH}
\subsection{Dynamical properties}
The local dynamics of Lorentz gases with infinite horizon, such as a
cubic lattice of spheres for $d\geq 2$ is somewhat more involved
due to orbits which are tangent to infinitely many scatterers.  In the
vicinity of such orbits are an infinite number of singularities
corresponding to orbits tangent to more and more remote scatterers.
However, the billiard map in two dimensions still satisfies exponential
decay of correlations~\cite{Chernov99}.

For the flow, results are given in~\cite{Melbourne09}: The decay of
correlations is known to be $O(1/t^{1-\epsilon})$ for general infinite
horizon Lorentz gases and explicitly $C/t$ for the standard example
of a square lattice of disks.  The space of functions considered does
not include position or velocity, however.  Extension to the $C/t$
result for arbitrary two dimensional Lorentz gases
is claimed in a later preprint~\cite{BM10}.

The situation for decay of correlations for infinite horizon Lorentz
gases in higher dimensions, even under reasonable assumptions, appears
to be open; see for example Ref.~\cite{BBT08}.  For the map it is likely
exponential, while for the flow it likely has the same asymptotic form
as $F(t)$, see Ref.~\cite{D12}.

\subsection{Transport}\label{s:DIH}
The two dimensional case is now relatively well understood, following a
number of approximate/numerical~\cite{ZGNR86,MM97,Bleher92} and
rigorous~\cite{Bleher92,SV07,CD09b} studies.  Recurrence and ergodicity
in the full space continue to hold, despite anomalous scaling.  Detailed
estimates for the recurrence properties (for example first return time
distribution) for a related random walk model may be found in
Ref.~\cite{Nandori11}.  The current is typically well defined if an external
field is not parallel to a horizon~\cite{CD09b}.  The long flights lead to
the other equations of  Sec.~\ref{s:diff} modified as follows:
\begin{description}
\item[Mean square displacement]
\begin{equation}
{\cal D}_{ij}=\lim_{t\to\infty}\frac{1}{2t\ln t}\langle\Delta_i\Delta_j\rangle
\end{equation}
where now ${\cal D}_{ij}$ is a superdiffusion matrix/tensor.
\item[Central limit theorem]
\begin{equation}
\frac{\Delta(t)}{\sqrt{t\ln t}}\Rightarrow {\cal N}(0,{\cal D}_{ij})
\end{equation}
where $\cal N$ is the multivariate normal distribution.
\item[Brownian motion]
\begin{equation}
\frac{\Delta(st)}{\sqrt{t\ln t}}\Rightarrow W(s)
\end{equation}
for $s\in[0,1]$, where $W$ is the standard $d$-dimensional Wiener process
with covariance matrix ${\cal D}_{ij}$.
\item[Local limit theorem]
Given an unbounded sequence of times $t_n$ and scatterers with displacements
$\Delta_n/t_n\to {\bf x}$ and Voronoi cells (excluding the scatterers
themselves) of volume $V_n$, the probability $P_n$ of reaching the cell at
time $t_n$ has the expected limit:
\begin{equation}
\frac{(t_n\ln t_n)^{d/2}P_n}{V_n}\to \phi({\bf x})
\end{equation}
where $\phi$ is the density function of ${\cal N}(0,{\cal D}_{ij})$.
\end{description}

The inconsistency with regard to factors of two compared with the normal
diffusion case was not noted until 2011; see Refs.~\cite{D12,BCD13}, where a
history and heuristic argument may be found.  Briefly, convergence in
distribution does not imply convergence of the moments, and in this case
as $t\to\infty$ the tails of the distribution decay in time while increasing
in extent so that contribution to the mean square displacement from the
tails does not decay, and in fact remains roughly equal to that from the
limiting normal distribution.  Recent results on convergence of moments in
fairly general dynamical systems (not including the infinite horizon Lorentz gas)
may be found in Ref.~\cite{MT12}.

Ref.~\cite{BCD13} specifically considers anomalous convergence of moments,
but in the context of billiards with cusps, that is, with the time to collision arbitrarily
small.  In this case the decay of correlations are algebraic for the map, but rapid
for the flow~\cite{BM08}.  As with the infinite horizon Lorentz gas (and also
the stadium billiard~\cite{BG06}) a nonstandard
(logarithmic) central limit theorem applies~\cite{BCD11}.

In contrast to the normal diffusion case (for example finite horizon),
the superdiffusion coefficient $\cal D$ can be expressed exactly in terms
of the geometry of the horizons (ie set of infinite orbits).  
Expressions are given for two dimensions
in Refs.~\cite{Bleher92,SV04,CD09b}.  In general we have (cf Sec.~\ref{s:hor}):
\begin{equation}
{\cal D}_{ij}=\frac{1}{1-{\cal P}}\frac{V_{d-1}}{S_{d-1}}
\sum_{H\in\mathbb{H}}\frac{w_H^2(\delta_{ij}-n_i(H)n_j(H))}{{\cal V}_H^\perp}
\end{equation}
if there is at least one non-incipient principal horizon, and for $d\geq 3$
subject to a conjecture that correlations decay more rapidly than $C/t$
restricted to trajectories with at least one collision~\cite{D12}.  Here,
$V_{d-1}$ is the volume of a $d-1$ dimensional ball, $w_H$ is the width of
the horizon (that is, one dimensional volume of $B_H$) and ${\bf n}(H)$ is
a unit vector parallel to $B_H$.

If the maximal horizon is not principal, we expect that correlations decay
as $1/t^{d-d_H}$, and thus from Eq.~(\ref{e:GK}) that the normal diffusion
coefficient exists (at least in terms of mean square displacement).  As
with finite horizon this is not accessible in closed form, but may be
approximated using correlated random walks~\cite{GNS11}.
As discussed in Ref.~\cite{CD09b}, if the horizon directions do not span the
full space, diffusion is anomalous in directions spanned by the horizons but normal 
in other directions.

\subsection{External fields}
Ref.~\cite{CD09b} also considers superdiffusion with a Gaussian thermostat,
finding  
\begin{equation}\label{e:JDI}
{\bf J}=\frac{1}{2}{\cal D}{\bf E}\ln|{\bf E}|+O({\bf E})
\end{equation} 
as long as the field is not parallel to a corridor (and the constant in
the error term may depend on the direction of $\bf E$).
The special case with corridors in only a single direction is also covered.
When the field is parallel to a corridor, motion in this direction is an
absorbing state, so all but a set of zero measure of initial conditions
achieve this.

For the Galton board with infinite horizon the scaling
$v\sim (t\ln t)^{1/3}$ was conjectured in Ref.~\cite{KR97}.

The thermostat-type model of Ref.~\cite{BR01} (see Sec.~\ref{s:weak})
was also considered in an infinite horizon regime.  As above, the
conductivity (current divided by field) was numerically observed
to be logarithmic in the field.

\section{Alternative limits}
Two limits that are required for a study of diffusion are $t\to\infty$ and $L\to\infty$
where $L$ is a length scale.  We have seen that for normal diffusion (for example finite
horizon Lorentz gases) these may be taken together, with $L=c\sqrt{t}$ to give an
appropriate limit law.  Similarly $L=c\sqrt{t\ln t}$ for the kind of superdiffusion found
with principal infinite horizons.  Other limits are also useful to consider, also involving
the scatterer radius $r$.
\subsection{Escape from finite domains}\label{s:esc}
The ``escape rate formalism'' originated with Gaspard and Nicolis~\cite{GN90}; see
also~\cite{GDG01,Gaspard05}.  Here we consider a finite horizon Lorentz gas on a finite
domain, say a square of side length $L$ (infinite domains such as a strip have also
been considered).  Scatterers from a periodic lattice are present inside
the square, but the region outside is empty (or else the particle is absorbed when it reaches
the boundary). Uniformly distributed initial conditions leak out, with a survival probability
$P(t)$ decaying exponentially in time $t$ due to the strongly chaotic dynamics.  The escape rate is
\begin{equation}
\gamma=-\lim_{t\to\infty}\frac{1}{t}\ln P(t)
\end{equation}
Results relating escape rates with Lyapunov exponents and entropy in a general setting and
for the two dimensional finite horizon Lorentz gas are proven in Ref.~\cite{DWY11}.  This
includes the ``escape rate formula,'' long known in simpler contexts~\cite{ER85,KG85}:
\begin{equation}
\gamma=\lambda-h_{KS}
\end{equation}
where the Lyapunov exponent and entropy refer to the natural invariant measure on the
non-escaping set.  Again, there is a Young formula for dimension~\cite{Young82,Gaspard05}
corresponding to Eq.~(\ref{e:DI}), although this does not appear to have been discussed
for the Lorentz gas in the recent rigorous literature
\begin{equation}
D_I=1+\frac{2h_{KS}}{\lambda}=3-\frac{2\gamma}{\lambda}
\end{equation}

We have taken the $t\to\infty$ limit at fixed $L$.  Now, for large $L$
we can compare with the hydrodynamic limit, Eq.~(\ref{e:hydro}) with the appropriate
absorbing (ie Dirichlet) boundary conditions.  For example, in the case of the
square $[0,L]^2$ and isotropic diffusion matrix $D_{ij}=D\delta_{ij}$, the density
\begin{equation}
\rho=e^{-\gamma t}\sin\frac{\pi x}{L}\sin\frac{\pi y}{L}
\end{equation}
is the lowest mode of the equation Eq.~(\ref{e:hydro}) if
\begin{equation}
\gamma=\frac{2D\pi^2}{L^2}
\end{equation}
Thus we can express the diffusion coefficient in terms of the escape rate and take
the limit $L\to\infty$.  Similar approaches can be made for other transport coefficients
such as viscosity~\cite{VG03}.  While not a practical method of computing
transport coefficients, it does not modify the equations of motion (unlike the Gaussian
thermostat in Sec.~\ref{s:gauss} above), and so is easier to justify from a physical
point of view.

In the infinite horizon case, little is known, though there is recent work where the
lattice is infinite and holes are located in the reduced space~\cite{Demers13}.

\begin{problem}
Quantify the time-dependence of the survival probability and size-dependence of the
escape rate for an open infinite horizon Lorentz gas.
\end{problem}

A boundary between two finite Lorentz gases with different parameters was recently
considered in Ref.~\cite{TY12}.  This work demonstrated the need for a careful
treatment of boundary conditions when considering hydrodynamic limits, namely
that the diffusion equation is not a complete description of the macroscopic system.

\subsection{The Boltzmann-Grad limit}\label{s:BG}
Here, the scatterer radius is taken to zero, but the spacing is also reduced so that the
mean free path remains fixed, as is useful in kinetic theory of low density gases, for
example the Boltzmann equation.   The distribution of path lengths has explicit formulae
available in two dimensions~\cite{Dahlqvist97,BZ07}.  It turns out that the linear
Boltzmann equation used by Lorentz for the random model (Sec.~\ref{s:random} below)
needs to be generalised, since for periodic models the limiting process has a kernel that
depends not only on the initial and final velocities at a single collision, but also the flight
time and velocity following a subsequent collision~\cite{BU09,CG08,MS08}.  The resulting
linear operator does not satisfy the semigroup property, despite being the zero radius limit
of operators that do.  

Marklof and Str\"omberggson extended this to arbitrary dimension;
technical proofs are found in Refs.~\cite{MS10,MS11a,MS11b}.  The tail of
the (closely related) free flight function in all dimensions agrees with the leading
term of Eq.~(\ref{e:Fsmall}) despite the differing manner in which the limit is
taken~\cite{D12}.  Marklof and T\'oth have recently proved a central limit theorem,
also in arbitary dimension~\cite{MT14}.  Their approach, which uses dynamics in the
space of lattices, has also led to interesting results in other fields, such as the
distribution of Frobenius numbers~\cite{Marklof10b}.  Readable reviews of the
work in this section are given in Refs.~\cite{Marklof14,Golse12}.
%Marklof10a

\section{Semi- and non- dispersing models}\label{s:nondis}
\subsection{Molecular dynamics}\label{s:MD}
\subsubsection{General discussion}
Much of the physical motivation for the periodic Lorentz gas is for
understanding molecular dynamics, models of many atoms moving under
Newton's laws, often using periodic boundary conditions~\cite{D00,EM08}.
These models shed light on theoretical issues,
such as the foundations of statistical mechanics, and the numerical simulations 
allow computation of how the bulk properties of materials depend on the
microscopic force laws and parameters such as energy and volume per particle.
The use of periodic boundary conditions avoids boundary effects, so bulk
properties can be estimated with fewer particles.

The Lorentz gas as we have discussed so far is equivalent to a two particle
system after transformation into (trivial) centre of mass motion and
the relative motion of the particles.  With three or more particles, there
is still equivalence with a high dimensional periodic billiard, however the
dispersing condition needs to be relaxed.  There are various definitions of
``semidispersing'' billiards in the literature; here we want to allow
cylindrical curvature, that is, positive curvature at all points, but not
in all directions.  Physically, when two particles collide, the outcome is
indifferent to the location and motion of the remaining particle(s).

The system of many hard particles has been a major motivation and stimulus for
the development of ergodic theory~\cite{Szasz00}.  Hyperbolicity is now known
for all hard ball systems~\cite{Simanyi02}.  Ergodicity is known when
$N=2$ (as above)~\cite{SC87}, $N=3$~\cite{KSS91}, $N=4$ for
$d\geq 3$~\cite{KSS92} and general $N\leq d$~\cite{Simanyi92a,Simanyi92b};
see also Ref.~\cite{BCST02}.  More recently ergodicity has been shown for
almost all parameters (masses and a single radius)~\cite{Simanyi04}, and
conditional on the Chernov-Sinai ansatz, the statement that almost every
singular orbit is hyperbolic~\cite{Simanyi09}.  Finally, there is a complete
proof in full generality~\cite{Simanyi13}.

While an impressive result, this does not spell the end of the subject~\cite{Szasz08}:
\begin{problem}
Is the system of hard balls in a hard box ergodic?
\end{problem}

\subsubsection{Infinite horizon effects}
The periodic boundary conditions usually lead to infinite horizon effects, that is, there
are trajectories in which one or more (usually all) of the particles never collide.  The
following is a previously unpublished study, mostly restricted to $N=3$ on a unit $2$-torus
and with all masses and radii equal and zero total momentum.

First note that if there are no collisions, the relative displacements of pairs of particles
are tracing out a lower dimensional affine space (including possibly a single point) 
that does not intersect the origin.  When $d=2$ this means that each relative velocity
lies in a rational direction.  If all relative velocities are parallel, they may be perturbed
parallel to this direction while remaining in the horizon, thus there is an $N-1$ dimensional
horizon (there are two constraints due to energy and parallel momentum conservation,
and one extra dimension from the flow direction).  The billiard itself is of dimension $2N-2$
(two components of momentum conservation), thus we expect a free flight function decaying
as $t^{-(N-1)}$, which is quite observable in $N=3$, leading to normal diffusion but
anomalous Burnett coefficients.  If relative velocities are not parallel, there is a lower
dimensional space of perturbations and hence a lower dimensional horizon.  Note that
on $\mathbb{T}^d$, it is also easy to see that for particles restricted to parallel hyperplanes,
the same decay of $t^{-(N-1)}$ ensues.

Parallel velocity directions are possible only if the particles are sufficiently small, that is, $2Nr<1$.
For a given radius we enumerate non-zero lattice vectors of length $L>(2Nr)^{-1}$ which correspond
to horizons (modulo reflection through the origin).  The space perpendicular to the horizon has
coordinates $x_i$, $1\leq i\leq N$ considered modulo $L^{-1}$.  However there is also a constraint
from the momentum conservation, which fixes the centre of mass: $\sum x_i=0.$  We define $x_i$
so that this constraint remains valid in a horizon, ie the $x_i$ do not translate when reaching the
boundaries of a fixed $L^{-1}$ interval; periodicity is taken account of by imposing that the maximum
$x_i-x_j$ is less than $L^{-1}$.

Specialising now to $N=3$ we construct orthonormal coordinates on the perpendicular space:
\[ \left(\begin{array}{c}x_1\\x_2\\x_3\end{array}\right)
=\frac{1}{6}\left(\begin{array}{cc}2\sqrt{6}&0\\ -\sqrt{6}&3\sqrt{2}\\ -\sqrt{6}&-3\sqrt{2}\end{array}\right)
\left(\begin{array}{c}\xi_1\\\xi_2\end{array}\right) \]
A fixed ordering of the particles $x_1>x_2>x_3>x_1-L^{-1}$ then corresponds to the equilateral triangle
$\xi_2>0$, $\sqrt{3}\xi_1-\xi_2>0$, $\sqrt{3}\xi_1+\xi_2<2/(L\sqrt{3})$ and similarly for the other
orderings.  The effect of finite radius is to tighten the inequalities further, $x_1>x_2+2r$ etc, and reduce
the size of the triangle.  Thus in contrast to the lower dimensional horizons associated with incipient horizons
discussed in Ref.~\cite{D12}, this non-principal horizon has a convex perpendicular space, for which the
visibility function is trivial.  For higher $N$, the corresponding perpendicular space is likewise a $N-1$
dimensional simplex, though not regular; for $N=4$ it is an isosceles tetrahedron.

We can now construct the various quantities appearing in Eq.~(\ref{e:F(t)}).  The latter formula assumed a
lattice of unit covolume; the covolume of the $\xi$-lattice is $\sqrt{3}$, and there is a similar factor from
the $y$-coordinates.  We need to divide the formula in Eq.~\ref{e:F(t)} by the covolume raised to the power
$1-{D_H}/d$, ie $\sqrt{3}$.  The double integral comes to $(L^{-1}-6r)^4/12$, taking into account the
finite radius of the balls.  We have $S_1=2\pi$ and $S_3=2\pi^2$.  Thus we find for a single horizon
\begin{equation}
F_H(t)=\frac{L^2(2\pi)(L^{-1}-6r)^4/12}{t^2(2\pi^2)\sqrt{3}(1-{\cal P})}
\end{equation}
The excluded volume is a slightly messy integral, giving (in the relevant region, $0\leq r\leq 1/6$)
\begin{equation}
{\cal P}=4\pi r^2(3-r^2(8\pi+3\sqrt{3}))
\end{equation}
For each primitive lattice vector there are six horizons, corresponding to the permutations of the particles,
however we divide by two, since opposite lattice vectors correspond to the same horizon.  This leads to
\begin{equation}\label{e:Fmd}
F(t)\sim\sum'_{{\bf l}\in\mathbb{Z}^2}\frac{L^2(L^{-1}-6r)^4}{t^24\pi\sqrt{3}(1-{\cal P})}
\end{equation}
where the sum is over primitive lattice vectors $\bf l$ of length $L$, for which $L^{-1}-6r$ is positive.

\begin{figure}
\centerline{\includegraphics[width=250pt]{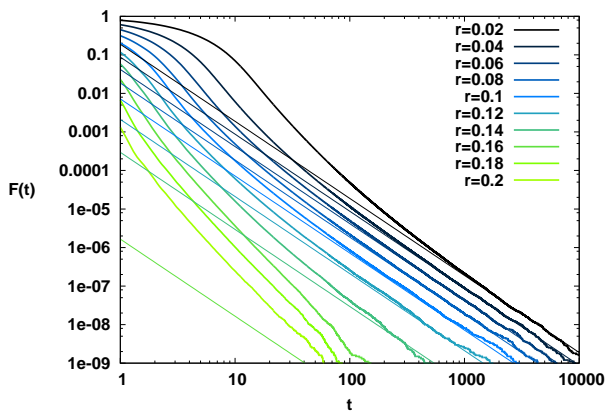}}
\vspace{-25pt}
\caption{Free flight function for three hard disks (thick curves)
together with predictions of Eq.~(\protect\ref{e:Fmd}) for $r<1/6$ (thin lines).\label{f:md}}
\end{figure}

Fig.~\ref{f:md} shows numerical simulations of $F(t)$ for various $r$, together with predictions from
Eq.~(\ref{e:Fmd}).  Note that for $r=0.16$ which is very close to $1/6$, the coefficient is very small
and would require greater times and sample sizes to observe.  Also, for $r>1/6$ there appear to be
$t^{-3}$ asymptotics, due to one-dimensional horizons, for example if $r<1/4$ is it possible to have
two particles following the same track with equal velocities and another particle in a parallel track moving
in the opposite direction.   There may be other contributions, however.

For $r\to 0$ we can extract the limiting behaviour using Mellin transforms as in Ref.~\cite{D12}.  The
result is
\begin{eqnarray}\nonumber
&&\lim_{t\to\infty}t^2 F(t)=\\\nonumber
&&\frac{\sqrt{3}}{\pi^2}\left(
-\ln r +\gamma+\ln\frac{\pi^{3/2}}{3\Gamma(1/4)^2}-\frac{\zeta'(2)}{\zeta(2)}-\frac{25}{12}\right)\\
&&+O(r^{3/2-\delta})
\end{eqnarray}
where $\gamma$ is the Euler constant and the correction term assumes the Riemann Hypothesis.

We can conclude the infinite horizon effects are definitely observable in the free flight function, decaying as
$t^{-(N-1)}$ for a configuration in which each particle moves parallel to a single lattice vector.  The coefficient
decreases with radius, and can be calculated explicitly for $N=3$.  These effects are not strong enough to
lead to an anomalous diffusion coefficient, but however lead to anomalous Burnett coefficients.
The main message, however is that simulations involving a few small particles have spurious long time
correlations due to periodic boundary conditions. 

\subsection{Moving scatterers}
Intermediate between the Lorentz gas and many-particle systems lie models with additional degrees of freedom.
We have already considered a few such models, the Nos\'e-Hoover thermostat in Sec.~\ref{s:weak} and
few-particle systems in Sec.~\ref{s:MD}.  A further class of models retains scatterers with fixed average
positions, normally on a periodic lattice, but which are moving in some manner.

\paragraph{Vibrating}
The most obvious effect of a moving boundary is in changing the speed of the particle; the collision law,
Eq.~(\ref{e:coll}) generalises to~\cite{BR11}
\begin{equation}
{\bf v}_+={\bf v}_--\frac{2{\bf n}}{\bf n\cdot n}({\bf n\cdot v_- - n\cdot u})
\end{equation}
where $\bf u$ is the velocity of the boundary.  This can lead to unbounded average particle speeds, often
termed ``Fermi acceleration''~\cite{Fermi49}. Refs.~\cite{LRA99,LCR08} discuss this in the context of both
stochastic and periodically moving scatterers for the finite horizon Lorentz gas.  In particular, the authors
of Ref.~\cite{LRA99} proposed (the ``LRA Conjecture'') that chaotic motion in the corresponding static
billiard was sufficient for Fermi acceleration.  More recently acceleration has been observed numerically
for the ellipse~\cite{LDS08}.  It is also known for a rectangle with a moving barrier~\cite{STR10} for
which the velocity growth is exponential.  Other examples and rigorous results for Fermi acceleration
in billiards are reviewed in Ref.~\cite{GRT12}.

In chaotic billiards the growth of velocity is typically proportional to the square root of the number of
collisions, hence linear in time, though slower rates have been observed
for ``breathing'' billiards that retain the same shape~\cite{BR11,Batistic13}.
Boltzmann equations and generalisations have been used to study the distribution of velocities, which
typically has an exponential rather than normal tail~\cite{JS93}. Recent work in this direction has
included periodically oscillating billiards~\cite{KDC12}, a Lorentz gas with stochastically moving scatterers~\cite{DK11,GPSM12}, and more general stochastic processes~\cite{KACK13}.

Infinite horizon effects have recently been considered for vibrating~\cite{KDPS12} and
pulsating~\cite{DL13} Lorentz models.  In this case there is a scenario of ``dynamically infinite horizon''
in which the billiard has infinite horizon only for part of the time.   Ref.~\cite{KDPS12} showed that
horizon effects led to power law correlations between non-interacting particles in a Lorentz channel,
while ref.~\cite{DL13} showed that these effects led to logarithmically enhanced ($v\sim t\ln t$)
Fermi acceleration.  In view of Sec.~\ref{s:poly} below

\begin{problem}
What rates of acceleration and diffusion are possible for time-dependent polygonal (Ehrenfest) models?
\end{problem}

\paragraph{Rotating}
A further generalisation is for each scatterer to have its own degree(s) of freedom.  A rotation-inspired
model with arbitrarily many degrees of freedom was studied in Ref.~\cite{RKN00}.  A model with
explicit rotating scatterers was proposed and used to study a number of transport phenomena in
Refs.~\cite{LLM03,SLL09}, and a mix of rotating and static scatterers was considered in Ref.~\cite{EY04}.
Thermal efficiency properties were studied using many internal degrees of freedom in Ref.~\cite{CMP08}.
In each of these models, the transfer of energy now permits normal heat conduction, with the dispersing
geometry as before used as a source of dynamical randomness.

\subsection{Flat points}\label{s:flat}
Another manner in which a Lorentz gas may become non-dispersing is the presence of points of zero curvature.
Such a model was considered in the finite horizon case in Ref.~\cite{CZ05} where it was shown that two points
with local graph $y=|x|^\beta$ (for $\beta>2$) forming a period two orbit leads to decay of correlations in
discrete time of roughly $n^{-(\beta+2)/(\beta-2)}$ rather than exponential.  Later~\cite{Zhang12} an
infinite horizon model was considered, containing an infinite trajectory tangent to a periodic sequence of such
flat points, leading to bounds on the free flight function and proofs of nonuniform hyperbolicity.

The following is study of a Lorentz gas with quartic flat points ($\beta=4$) that has not been previously
published. It is similar to Ref.~\cite{CZ05} in that the horizon is finite, however the flat points lead to
translating periodic orbits, which enhance the rate of diffusion.    We will see that the quartic flat point is
just sufficient to make the fourth order Burnett coefficient diverge logarithmically, leading to anomalous
convergence effects.
 
\begin{figure}
\centerline{\raisebox{80pt}{$y$}\hspace{-20pt}\includegraphics[width=250pt]{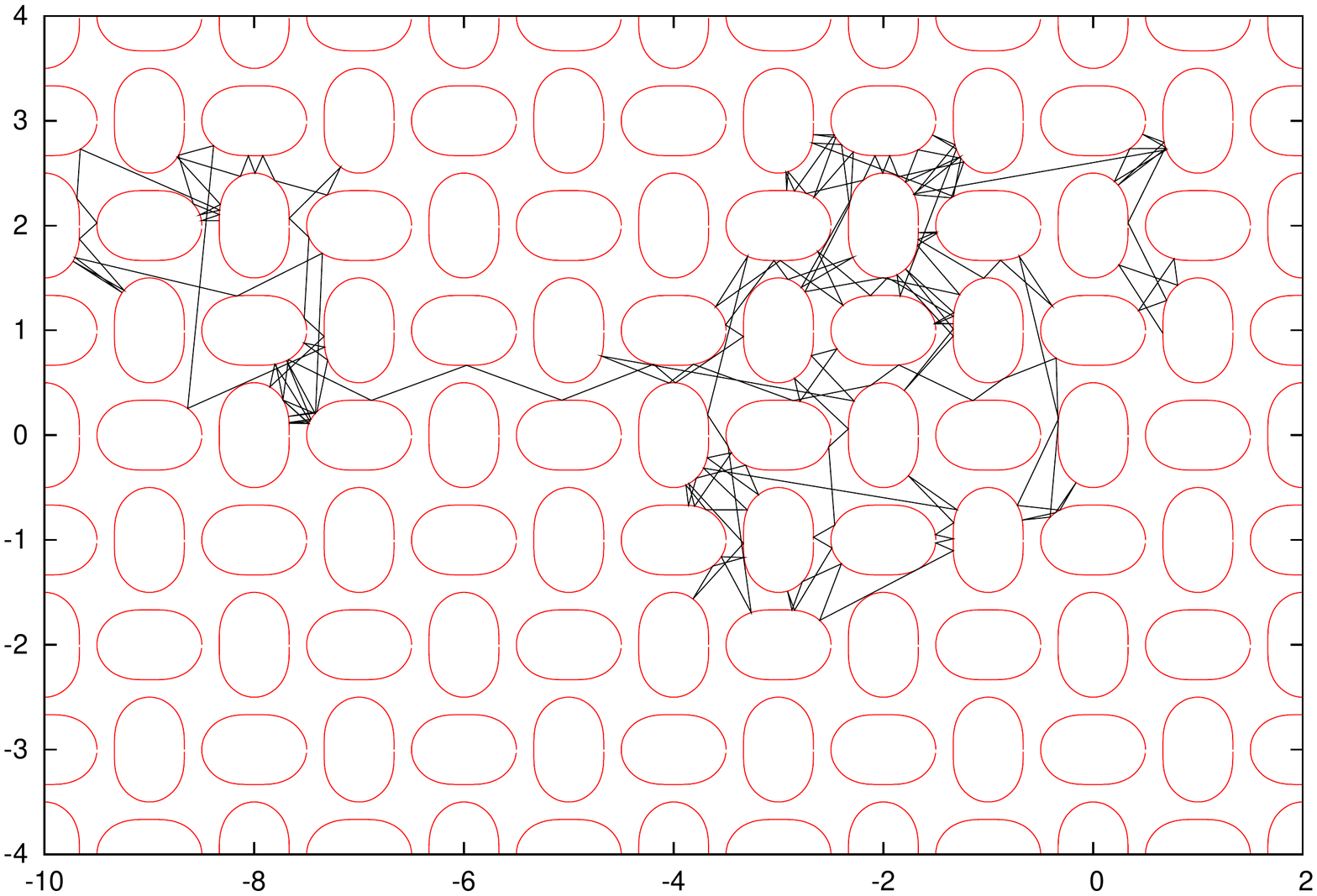}}
\vspace{-20pt}
\centerline{$x$}
\caption{A trajectory in the oval Lorentz gas\label{f:traj}}
\end{figure}

The scatterers are ovals defined in local polar coordinates $(r,\phi)$ by
\begin{equation}
r=\frac{5-(-1)^{I+J}\cos(2\phi)}{12}
\end{equation}
thus having semimajor axis $1/2$ and semiminor axis $1/3$.  The centres of the scatterers are
located at points of the integer lattice, $(I,J)\in\mathbb{Z}^2$.  The effect of the sign is
to rotate scatterers at odd points by $\pi/2$, a configuration with finite horizon
and no scatterers touching, see Fig.~\ref{f:traj}.  

The symmetries are those of a square, reflection in both axes and in $y=x$, the latter with a
spatial translation.  Thus the nonzero cumulants up to order four are $Q_{20}=Q_{02}=M_{20}$,
$Q_{40}=Q_{04}=M_{40}-3M_{20}^2$ and $Q_{22}=M_{22}-M_{20}^2$.

Each scatterer has an area $1-|S|=17\pi/96\approx 0.556324$, so that the area available
to the billiard particle in each unit cell is $|S|$.  The perimeter is given by the elliptic integral
\begin{equation}
|\partial S|=\frac{\sqrt{5}}{3}\int_{-\infty}^\infty
\frac{\sqrt{2t^4+2t^2+1}}{(t^2+1)^2}dt
\approx 2.72244
\end{equation}

Now consider the central scatterer ($I=J=0$).  Near the point $\phi=0$ we find
\begin{equation}
x=x_0-\kappa y^4+\ldots
\end{equation}
where $x_0=1/3$ and $\kappa=81/8$.  Thus the curvature at this point is zero, and it has a
quartic shape, which is generic for analytic zero curvature (``flat'') points.  Each flat point can
reach exactly two other flat points by a free flight and belongs to exactly one marginally unstable
translating orbit (modulo time reversal).  These orbits translate either in the horizontal or vertical directions, for example the orbit from $(1/3,0)$ reaches the flat point $(2/3,1)$ and then reflects
to reach $(1/3,2)$.  Between each pair of flat points it translates one unit in its overall direction
of motion (here the $y$ direction), taking a time $\tau=\sqrt{10}/3$.  The angle of incidence is $\theta_0=\arctan 3\approx 72^\circ$.  A horizontal version of this orbit can be seen for several
collisions in Fig.~\ref{f:traj}.

Now we perturb the translating orbit.  The displacement from the flat point is $y$ (taken mod 1),
while the direction relative to the $x$-axis is $\theta_0+\theta$ so that both $y$ and $\theta$ are
small.  The approximate collision map is
\begin{equation}\label{e:disc}
y_{n+1}=y_n+\eta\theta_n,\qquad
\theta_{n+1}=\theta_n+\psi y_{n+1}^3
\end{equation}
For initial conditions $\theta\approx y^2$ which we will discover are typical, the relative
increments of each variable are small, so we can use a continuum approximation
\begin{equation}\label{e:continuum}
\frac{dy}{dn}=\eta\theta,\qquad
\frac{d\theta}{dn}=\psi y^3
\end{equation}
Here, $\eta=\delta x\sec^2\theta_0=10/3$, where $\delta x=1/3$ is the $x$ displacement of a single
flight, and $\psi y^3=-2dx/dy=8\kappa y^3=81y^3$. Thus we have
\begin{equation}\label{e:ODE}
\frac{d^2y}{dn^2}=\eta\psi y^3\qquad \Rightarrow \qquad
\left(\frac{dy}{dn}\right)^2=\frac{\eta\psi y^4}{2}+A
\end{equation}
where $A$ is a constant of integration, determined from the initial conditions.  To reach the
translating orbit itself we need $A=0$, which gives separatrix solutions
\begin{equation}
y=\pm\sqrt{\frac{2}{\eta\psi}}n^{-1}, \qquad
\theta=\mp \sqrt{\frac{2}{\eta^3\psi}}n^{-2}
\end{equation}
showing that the scaling $\theta\approx y^2$ is generally valid approaching the marginal orbit.
A second constant is omitted here as it just translates the collision time $n$.

\begin{figure}
\centerline{\raisebox{140pt}{$\theta$}\hspace{-20pt}\includegraphics[width=200pt]{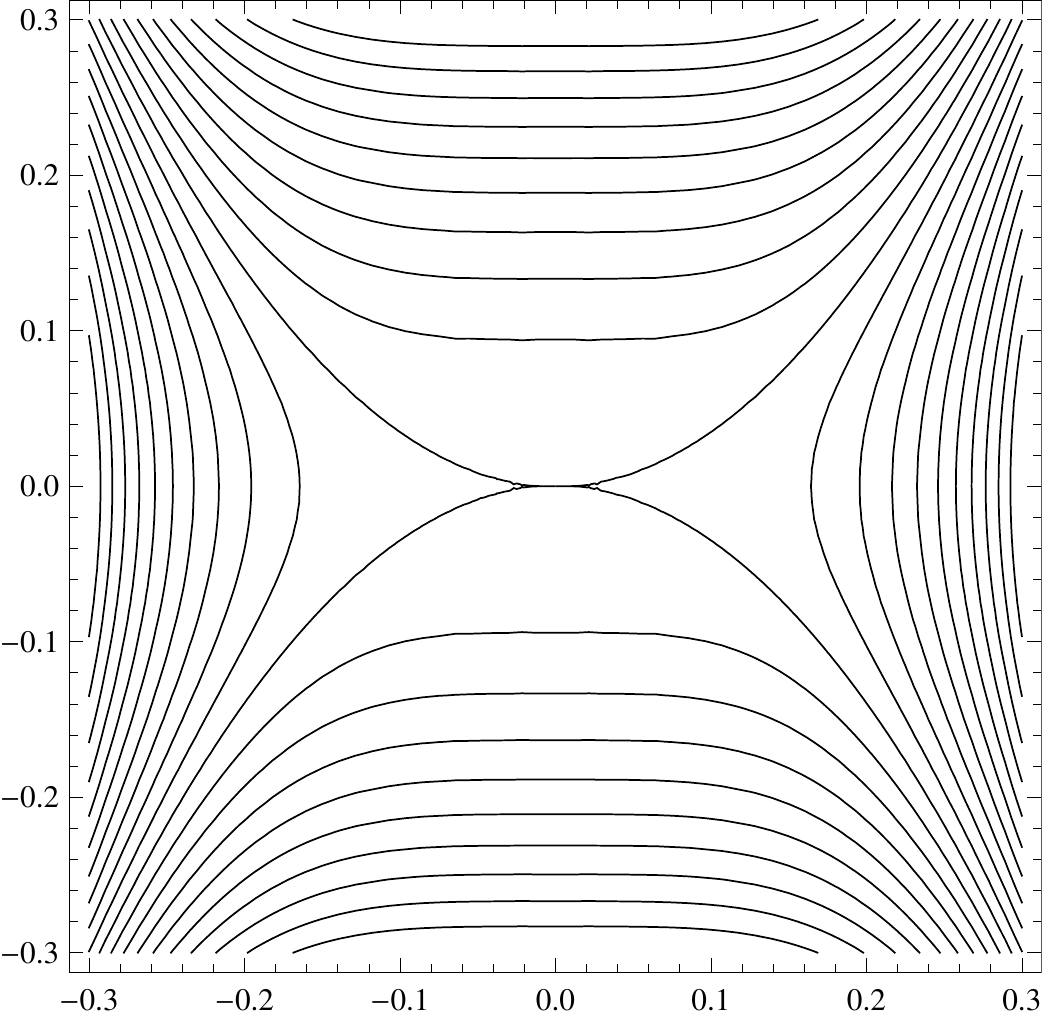}}
\vspace{0pt}
\centerline{$y$}
\caption{Orbits of the flow approximation, moving right at the top and
left at the bottom.\label{f:orbit}}
\end{figure}

For general $A$ we can combine the first part of Eq.~(\ref{e:continuum}) with the second
part of Eq.~(\ref{e:ODE}) to give an equation for the orbits
\begin{equation}\label{e:orbit}
\eta^2\theta^2=\frac{\eta\psi}{2}y^4+A
\end{equation}
which is plotted in Fig.~\ref{f:orbit}.  The flow direction is in from the top left and
bottom right, and out to the top right and bottom left.  

Trajectories which remain near the marginal point for a long time have $A$ close to zero.
Let us fix the vicinity of the marginal point as the interval $y\in[-Y,Y]$ for some
arbitrary small constant $Y$.  The traversal time $N(A)$ is then the number of iterations needed
to move from $y=-Y$ to $y=Y$ above the separatrix for $A>0$, or back to $y=-Y$ to the left
of the separatrix for $A<0$.  For $A>0$ we have
\begin{eqnarray}
N(A)&=&\int_{-Y}^{Y}\frac{dn}{dy}dy\nonumber\\
&=&\sqrt{\frac{2}{\eta\psi}}\int_{-Y}^{Y}\frac{dy}{\sqrt{y^4+2A/(\eta\psi)}}\nonumber\\
&=&\frac{\Gamma(1/4)^2}{(8\pi^2\eta\psi A)^{1/4}}+O(Y^{-1})
\end{eqnarray}
For $A<0$ we have
\begin{eqnarray}
N(A)&=&\int_{-\theta(Y)}^{\theta(Y)}\frac{dn}{d\theta}d\theta\nonumber\\
&=&(8\eta^3\psi)^{-1/4}\int_{-\theta(Y)}^{\theta(Y)}\frac{d\theta}{(\theta^2-A/\eta^2)^{3/4}}\nonumber\\
&=&\frac{\Gamma(1/4)^2}{(32\pi^2\eta\psi|A|)^{1/4}}+O(Y^{-1})
\end{eqnarray}
Where $\theta(Y)=\sqrt{(\psi Y^4)/(2\eta)+A}$ and both integrals were done with Mathematica.  Note
that the order of the relevant limits is $A\to 0$ at fixed $Y$ giving the long flight
behaviour, followed by $Y\to 0$.  Thus we can use $Y=\infty$ in the above integrals.  Physically,
for very long flights, almost all the collisions are very close to the marginal point, so the
size of the considered region becomes irrelevant.

We can now calculate the asymptotic coefficient of the probability of a long flight.  We cut a long
billiard trajectory into $M$ segments with displacement ${\bf x}_i$, $i=1\ldots M$ and continuous
time $t_i$.  Each segment is either a single collision, or a flight following the marginal orbit
for some time, so that correlations are expected to decay exponentially in the number of segments.
The continuous time is $t_i=|{\bf x}_i|$ for single collisions and $t_i\approx \tau|{\bf x}_i|$ for
very long segments; recall that $\tau=\sqrt{10}/3$ in the present example.  The total displacement and time are thus
\begin{equation}
\Delta=\sum_i{\bf x}_i,\qquad T=\sum_i t_i\sim \bar{t}M
\end{equation}
with an unknown (but ultimately irrelevant) constant $\bar{t}\approx 1$.

The anomalous behaviour arises from the long tail in the distribution of ${\bf x}_i$.  Its density
function $p({\bf x})$ is concentrated mostly around the origin, but has tails along the $x$ and $y$
axes due to the marginal orbits in these directions.  Each long trajectory of at least $N$
collisions near the marginal orbit enters the region of the marginal orbit $|y|=Y$ (or corresponding expression involving $x$) exactly once if $N$ is sufficiently large.  From the above calculation, the first collision lies in the region
\begin{equation}\label{e:Aint}
-\frac{C}{N^4}<A<\frac{4C}{N^4}
\end{equation}
where $C=\Gamma(1/4)^8/(32\pi^2\eta\psi)=\Gamma(1/4)^8/(8640\pi^2)\approx 0.350135$.  This region
corresponds to intervals
\begin{equation}
\delta\theta=\frac{5C}{\sqrt{2\eta^3\psi}}\frac{1}{Y^2N^4},\qquad
\delta y=\sqrt{\frac{\eta\psi}{2}}Y^2
\end{equation}
where the first expression is found by combining Eqs.~(\ref{e:orbit},\ref{e:Aint}) for small $A$,
and the second is just an approximation for the change in $y$ due to a collision at $|y|=Y$,
following from a combination of Eqs.~(\ref{e:disc},\ref{e:orbit}), again for small $A$.  

The measure of one of the above regions with respect to the equilibrium measure of the billiard map in the torus
is
\begin{equation}
\frac{\cos\theta_0\delta\theta\delta y}{2|\partial S|}=\frac{\sqrt{10}C}{8\eta|\partial S|N^4}
\end{equation}
Thus, the expected number of excursions of more than $N$ collisions in a trajectory of
total length $T$ is given by
\begin{equation}
\frac{\sqrt{10}CT}{\eta\bar\tau|\partial S|N^4}=\frac{\sqrt{10}\Gamma(\frac14)^8T}
{300\pi^3(96-17\pi)N^4}
\end{equation}
where the factor of 8 takes account of the four orbit directions (up, down, left, right) and the
two entry points $y=\pm Y$ (or the same with $x$) of the marginal orbit.  The mean free
path between collisions is $\bar\tau=\pi |S|/|\partial S|$ as for any 2D billiard table.  Strictly
speaking this is for the entry to the marginal point; if $N>T/\tau$ (very unlikely), clearly
the collisions cannot all take place in the interval.  A typical interval will have many
excursions, with the largest almost always $O(T^{1/4})$ according to the above formula.

This gives the tail of the density function $p({\bf x})$ as
\begin{equation}
\int_Y^\infty dy \int_{-\infty}^\infty dx p(x,y) \sim \frac{D\bar{t}}{Y^4}\qquad Y\to\infty
\end{equation}
where
\begin{equation}
D=\frac{\sqrt{10}\Gamma(\frac{1}{4})^8}
{1200\pi^3(96-17\pi)}\approx 0.0595772
\end{equation}
is an explicit constant.  The factor of 4 appears since this is one of four tails,
$T=\bar{t}M$ as above, and the displacement at each collision $\delta y=1$ so $Y=N$.

The above asymptotic implies that fourth moments of $p({\bf x})$ diverge.  A finite sample will
have
\begin{equation}
\frac{1}{M}\sum_i y_i^4\approx 2\int_{E_1}^{E_2 M^{1/4}} y^4\frac{4D\bar{t}}{y^5}dy \sim 2D\bar{t}\ln M
\end{equation}
where $E_1$ and $E_2$ are constants of order unity, and the $M^{1/4}$ gives the scale of the
largest excursion.  The factor of 2 gives the two tails in the $y$ direction contributing to
this moment.  The sum of $x_i^4$ is equivalent, while the mixed even fourth moment $x^2y^2$
and second order even moments $x^2$ and $y^2$ are small in all tails and yield finite constants.
The odd moments cancel by symmetry, and are of size $M^{-1/2}$, at least up to fourth order.

\begin{figure}
\centerline{\includegraphics[width=250pt]{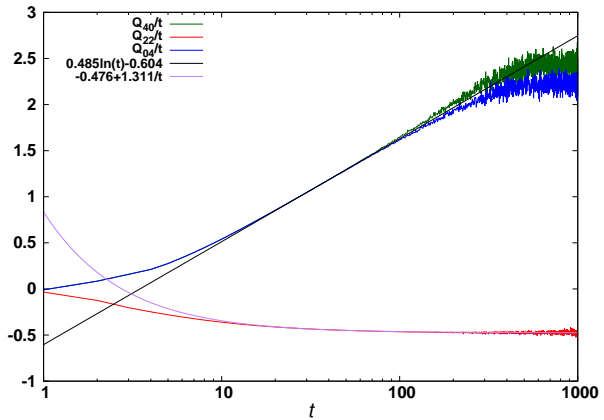}}
\vspace{-30pt}
%\centerline{$t$}
\caption{Logarithmic divergence of Burnett coefficients, using a trajectory of length
$7.41\times 10^{10}$ time units split into segments of length $t$, and relevant fits.\label{f:Burnett}}
\end{figure}

Referring back to Sec.~\ref{s:Burnett} we have the Burnett coefficient
\begin{equation}
24B_{1111}=\lim_{T\to\infty}\frac{Q_{40}}{T}
\end{equation}
Here, $\langle\rangle$ is an ensemble at fixed $T$, so that the fourth moment is finite (unlike for fixed $M$).
A typical trajectory has $M\approx T/\bar{t}$.  Expanding the $\Delta$ terms, assuming the $x_i$ are
independent and have zero odd moments gives
\begin{eqnarray}
24B_{1111}(T,1)&=&\frac{1}{T}\sum_i\left[\langle x_i^4\rangle-3\langle x_i^2\rangle^2\right]\nonumber\\
&=&2D\ln T+O(1)
\end{eqnarray}
This is the typical Burnett coefficient found for a single trajectory of length $T$.  The arguments are
the time and the sample size over which it is averaged.  This is not however its expected value, which includes the
distribution of $p(x,y)$ up to the maximum length $T$.
Truncating at the maximum $T$ rather than the typical $T^{1/4}$ gives a further factor of 4 for the
Burnett coefficient as usually defined, that is, averaged over an arbitrarily large sized sample
\begin{equation}
24B_{1111}(T,\infty)=\frac{\langle \Delta_x^4\rangle-3\langle \Delta_x^2\rangle^2}{T}\sim 8D\ln T
\end{equation}
The constant $8D$ comes to about $0.476618$, so agreeing with the numerically fitted 0.485 in
Fig.~\ref{f:Burnett}, and exhibiting an anomaly similar to the anomalous convergence of the
second moment observed in the diffusion case~\ref{s:DIH}: The logarithmic Burnett coefficient is
a factor of four greater than its typical value estimated from a single trajectory of the same length.
In other words, while a typical trajectory of fixed continuous time $T$ has a maximum excursion
of order $T^{1/4}$ in the limit $T\to\infty$, the fourth moment picks up the full support up to $T$.
The fixed $M$ moment is likewise infinite, as it picks up contributions from arbitrarily long excursions.

For a sample size $T^\alpha$, there are $M=T^{\alpha+1}$ excursions, so that the largest is likely to be
of order $(T)^{(\alpha+1)/4}$ leading to an estimate $2D(\alpha+1)\ln T$ where this is less than $8D\ln T$.
Some decrease is indeed observed to the right of Fig.~\ref{f:Burnett}, but also random fluctuations.  More
properly, with high probability we expect for a sequence of trajectories of increasing length
\begin{equation}
\lim_{T\to\infty}\frac{24 B_{1111}(T,T^\alpha)}{\ln T}= 2D\max(\alpha+1,4)
\end{equation}
Of course, this would require careful arguments and estimates to justify use of the various assumptions,
and it may require unreasonably long times in practice before the $\ln T$ damps the nonleading contributions.

The long flights are only in the coordinate directions, and do not lead to diverging second moments.  Thus
\begin{equation}
24B_{1122}=\lim_{T\to\infty}\frac{Q_{22}}{T}
\end{equation}
approaches a finite limit as in the figure.  It would be interesting to see if the logarithmic Burnett coefficients
and anomalous convergence lead to corresponding effects in application areas, from local limit theorems to
molecular dynamics.

\subsection{Polygonal scatterers}\label{s:poly}
The $\beta\to\infty$ limit of the infinite horizon model with flat points considered in Ref.~\cite{Zhang12}
is that of a square.  If there are no points with non-zero curvature, the Lyapunov exponents and entropy
are zero, and dynamics is dominated by the remaining singularities, the corner points.  Polygonal billiards
thus have properties very different to the dispersing case, and have been under very active investigation
recently.

For billiards inside polygons, angles of the form $\pi/n$ are removable singularities.
Thus the polygons with angles $(\pi/2,\pi/2,\pi/2,\pi/2)$, $(\pi/2,\pi/4,\pi/4)$,
$(\pi/2,\pi/3,\pi/6)$ and $(\pi/3,\pi/3,\pi/3)$ are completely regular, as are polyhedra associated with
Coxeter groups~\cite{PS98}.  Other polygons with angles a rational multiple of $\pi$ can be mapped
to translation surfaces of finite genus, and any trajectory can have only a finite number of velocity
directions.  These have been widely studied for the last decade, with many results surveyed
in Ref.~\cite{Gutkin12}.  The flow is uniquely ergodic in almost all directions~\cite{KMS86},
weak mixing in almost all directions for most regular polygons~\cite{AD13,Ferenczi13}, and rational
polygonal billiards are never strong mixing~\cite{Katok80}.  Irrational angles are much more difficult to
study. In some cases there is numerical evidence of strong mixing~\cite{CP99} which is widely
disbelieved but not disproved, while in others, non-ergodicity~\cite{WCP13}.  A well known
problem is~\cite{Schwartz09}

\begin{problem}
Do all triangular billiards have at least one periodic orbit?
\end{problem}

An extended billiard with square scatterers is called the Ehrenfest wind-tree model, proposed in 1912 by P.
and T. Ehrenfest~\cite{Ehrenfest}.  The original model, like the original Lorentz model, had dilute randomly
placed scatterers (``trees''), which were parallel, for example with their diagonals along the axes.  The
particle (``wind'') in the original model had an angle of incidence of $\pi/4$ at each collision, moving always
parallel to the $x$ or $y$ axis.  The corresponding three-dimensional model (with parallel rhombic
dodecahedra) seems never to have been studied, however the other conditions have been relaxed, allowing
other particle directions, non-parallel orientations, and other polygonal scatterers.   As with the Lorentz gas,
we first consider periodic configurations.

The first rigorous study of the periodic wind-tree was Ref.~\cite{HW80}, describing orbits with angle of
incidence $\pi/4$ as above, for which the dynamics reduces to that of a rotation.
For rectangular scatterers of size $((1+\alpha)/4,(1-\alpha)/4$, rational
$\alpha$ leads to orbits periodic in then reduced space (hence periodic or translating in the full space).
For irrational $\alpha$ information about the orbit can be obtained using the arithmetic properties (specifically
the continued fraction expansion) of $\alpha$ to obtain a logarithmically diverging sequence of points on
the trajectory, hence showing that it is unbounded~\cite{HW80}.

More general directions and models require the study of more general interval exchange transformations
than rotations, so that the next major result did not come until ref.~\cite{HLT11}.  Here it was shown that
for rectangles with rational lengths that (in lowest form) have odd numerator and even denominator, there
is a dense set of rational directions for which the dynamics is periodic, and that for almost all directions the
dynamics is recurrent.  Also, for rectangles with even numerator and odd denominator, there is a dense set
of rational directions for which no trajectory is periodic and almost all directions have a logarithmically
diverging sequence of points.  Thus for generic parameter values (in a topological sense), the dynamics
is recurrent, has a dense set of periodic points, and (at least) logarithmically divergent trajectories for
almost all directions.

Despite the recurrence results, almost all wind-tree and similar models are non-ergodic in almost all
directions~\cite{FU13} (in contrast to the finite and infinite horizon Lorentz gases above).
Finally, Ref.~\cite{DHL14} gives a detailed calculation using theory for
translation surfaces developed in recent years showing that almost all wind-trees and directions have
$\limsup\ln|\Delta|/\ln t=2/3$; presumably this is true for typical displacements as well,
though it is almost certainly too much to expect a limiting distribution.  

Diffusion in a polygonal honeycomb lattice was considered in Ref.~\cite{SS06}  The numerical
simulations show $\langle \Delta^2\rangle\sim t^{1.72}$ with an anisotropic distribution,
due to long flights in six equally spaced directions~\cite{SS06}.

A number of authors have performed numerical simulations for polygonal channels, that is,
a two dimensional geometry confined between parallel walls and periodic in the direction
parallel to the walls.  Variables include whether the horizon is finite, whether scatterers
are parallel, whether angles are rational or irrational multiples of $\pi$.  See for example
Refs.~\cite{ARV04,SL06,JR06}.  The observed diffusion included normal and anomalous with
various exponents, but a general theory appears to be absent.  

\begin{problem}
Classify diffusive regimes for polygonal channels.
\end{problem}

A deceptively simple example of a Lorentz channel, a ``barrier billiard'' consists of two parallel walls
with periodic infinitely thin spikes protruding perpendicular to one or both walls; models of this type were
considered in Ref.~\cite{Zwanzig83,HM90}.  For small spikes in one of the walls, this is
a retro-reflector, reversing almost all incoming trajectories~\cite{BKMP11}; it is also one of the
models shown to be non-ergodic in almost all directions in Ref.~\cite{FU13}.

An external field and Gaussian thermostat has also been considered. This leads to transient
behaviour followed by stable periodic orbits, for a rhombus wind-tree~\cite{LRB00,BR09} and polygonal 
channels~\cite{JR06}.  This work including a generalisation to finite particle size, provides insight
into anomalous diffusion phenomena in nanopores~\cite{JBR08,Bianca10}.  For the polygonal version
of the thermostat-type model of Ref.~\cite{BR01} (see Sec.~\ref{s:weak} above) collapse onto
a periodic orbit was observed, but with zero current.

In summary, there are a number of important results for some classes of polygonal scatterers where
all boundaries are in rational directions, however the case of irrational directions eludes understanding.
There are very many difficult remaining open problems. 

\section{Aperiodic models}\label{s:aper}
\subsection{Quasiperiodic models}
 We now consider aperiodic models, again
assuming the dispersing property except where indicated.
A quasiperiodic scatterer arrangement is a non-periodic model where the scatterer positions are
obtained by the cut-and-project method, that is, taking the intersection of a periodic lattice with
an infinite slab (more generally $\mathbb{R}^d\times S$ with $S$ a compact set)
at some irrational orientation, and projecting transverse to the infinite direction(s) of
the slab to obtain a non-periodic set in a lower dimensional space.  Note that other and more general
definitions are possible, for example using substitutions, tilings or separated nets~\cite{Solomon11}.
The transport properties of quasiperiodic Lorentz gases were posed as an open problem in
Ref.~\cite{Szasz08}, although quasiperiodic soft potentials had been investigated for some
time~\cite{DS75,ZSCC89}.

Using the cut-and-project method, it is possible to reduce the problem to
that of a periodic billiard in a higher dimension~\cite{KS13}.
This permits a natural probability measure for initial conditions, and
furthermore allows identification of infinite horizon channels (for relatively small
scatterers), which could be analysed as in the periodic case.  According
to the numerical simulations, diffusion is normal for finite horizon, slightly
superdiffusive when there is an infinite horizon, and slightly subdiffusive where the
scatterers can overlap.

In the Boltzmann-Grad limit, the free path length is numerically found to be algebraic as in the
periodic case~\cite{Wennberg12}, and indeed recent methods used to study the periodic Lorentz
gas (Sec.~\ref{s:BG}) can be applied here also~\cite{MS13a}, also for the
union of periodic lattices~\cite{MS13b}.  

\subsection{Local perturbations}
A periodic Lorentz gas may be locally perturbed, either by changing a finite number
of scatterers, imposing a local external field, or imposing a line that the particle
reflects from resulting in motion in a half-plane.  Ref~\cite{DSV09} shows that
convergence to Brownian motion with the same diffusion matrix still holds,
with differing boundary conditions where appropriate.  Refs.~\cite{PS10,Nandori11} made
the first steps to extending this to the infinite horizon case by showing analogous
behaviour for random walks with unbounded jumps, including with a $\sqrt{t\ln t}$
scaling.

\subsection{Decimation and Lorentz tubes}
Recurrence is generic (in a topological sense) for Lorentz and wind-tree (parallel rectangle) models
obtained by randomly deleting scatterers subject to a locally finite horizon condition~\cite{Troubetzkoy10}.
In a similar vein, a Lorentz tube is a one-dimensional lattice of cells, where the contents of each cell is
chosen randomly from a set of dispersing scatterer configurations.  Note that the problem of defining
the measure for the initial condition of the particle on an infinite space is circumvented: Place the particle
in the central cell and choose the scatterer configuration according to the specified distribution.
Lorentz tubes in two~\cite{CLS10} and higher~\cite{SLDC11} dimension with finite horizon are all
hyperbolic and almost all are recurrent, ergodic, and K-mixing.  The same properties hold in two
dimensions when the finite horizon condition is relaxed~\cite{LT11}.

For Lorentz gases where the scatterer is randomly changed each time the particle enters a cell,
stronger properties (vector almost sure invariance principle) may be shown~\cite{Stenlund12}.

\subsection{Limiting random models}\label{s:random}
Finally, we consider models in which the scatterers are placed randomly without reference to an underlying
lattice.  Progress has been made mostly for the low density (Boltzmann-Grad) limit so that to a first
approximation we may neglect overlapping scatterers and recollisions (ie collision with the same scatterer
more than once in a short time), that is, assume that the scatterer locations are a Poisson process.
Lorentz derived a linear Boltzmann equation (linear since the only one particle moves),
which was subsequently the subject of more rigorous studies~\cite{Gallavotti69,Spohn78}.
In particular, the linear Boltzmann equation holds when the scatterer density converges in probability
to its mean (so, not necessarily a Poisson distribution) and any soft potential has finite
range~\cite{Spohn78}. The Boltzmann-Grad and other related limits and many particle
models were reviewed in Ref.~\cite{Spohn80}, where the above results are restated for $d\geq 2$.
The Boltzmann equation for the Poisson-distributed two dimensional Lorentz gas is shown for
typical configurations in Ref.~\cite{BBS83}.

As with the Lorentz tubes, the initial position of the particle may be chosen as the origin, with scatterer
positions chosen randomly.  This ensemble may be used to define averages and correlation functions as
usual.  The random Lorentz gas exhibits power law decay of correlations, as $t^{-d/2-1}$ according
to low density kinetic theory.  As with similar behaviour for the multi-particle fluids for which it is a
prototype, this came as a surprise in the 1960s; these ``long time tails'' which lead to anomalous and
non-analytic behaviour (typically logarithmic terms) in transport coefficients.  So, we expect the diffusion
coefficient to exist but each Burnett coefficient only in sufficiently high dimension.  A detailed history
and discussion of these results may be found in Ref.~\cite{vB82}.

Kinetic theory methods have more recently been applied to the calculation of other dynamical
properties in dilute random Lorentz gases, including the Lyapunov exponents at equilibrium~\cite{vBD95,vBD96,WvB04,KPvB06}, with field and thermostat~\cite{vBDCPD96,MvB04},
and with open boundary conditions~\cite{vBLD00,vBM05}. This approach was also extended
to many particle systems~\cite{vZvBD98,WvB11}.  Again, logarithmic terms abound.

An applied magnetic field is considered in Refs.~\cite{BMHH97,KS98,BHPH01}.  The magnetic field bends
the trajectories into a circle, making recollisions likely, hence requiring a generalisation of the Boltzmann
equation.  Adding an additional weak electric field naturally causes a drift with a component perpendicular
to the fields, however a scatterer may also cause the particle to be trapped~\cite{BHHP96}. Current and
diffusion may be analysed, generalising the case without a magnetic field~\cite{PHH97}.

Weak coupling limit models, in which the particle is deflected only slightly when it reaches a scatterer,
were also reviewed in Ref.~\cite{Spohn80}. There has been recent progress in showing convergence
to the heat equation~\cite{BNP13}, also with a logarithmic correction.

For the random Ehrenfest model (parallel square scatterers, particle making collision angle $\pi/4$) kinetic
theory and numerical simulation show normal diffusion for non-overlapping scatterers but sub-diffusion if
the scatterers are allowed to overlap~\cite{HC69,WL71,vBH72}.  Many of the above theorems for the
Lorentz gas require only a smooth differential cross-section function and hence apply.  However, it is
noted in Ref.~\cite{Hauge74} that an Ehrenfest-like model with parallel crosses and incidence angle of
$\pi/4$ has a finite probability for the particle to immediately return to the previous scatterer, and hence
exhibits abnormal behaviour.

\subsection{Fixed random models}\label{s:fixed}
The case of randomly placed scatterers of fixed size has also been widely considered theoretically and numerically. 
For overlapping scatterers at high density there is a percolation transition, at which the diffusion coefficient
goes to zero with certain critical exponents and beyond which motion is localised~\cite{MM85}. More recent
discussion of these phenomena and simulations for the overlapping model in two and three dimensions
may be found in Ref.~\cite{HF07}. 

Random non-overlapping scatterers arise naturally for a mixture of small light particles and large heavy
particles in the limit of infinite mass and size ratios.   However, this model has resisted rigorous
results so far.  Even exhibiting the limit of a Poisson distribution of scatterers conditional on them
non-overlapping does not appear to have been attempted, although it is the hard potential limit of standard
results on the thermodynamic limit of systems with soft potentials; see for example Ref.~\cite{Uffink06}.

In numerical simulations at high density, it can be difficult to find a non-overlapping configuration directly; one
approach is to start with a periodic lattice, apply random velocities to all scatterers (as in a full molecular
dynamics simulation) and await relaxation to equilibrium.  Based on the low density results above, one
would again expect normal diffusion but anomalous Burnett coefficients for the random non-overlapping
model, and this is what is found.  Correlations are found numerically to decay at the same rate as
predicted in the low density limit after a time which increases with the density~\cite{AA83}.  

Making a diffusive scaling $L\sim \sqrt{t}$, numerical simulations of the non-overlapping model exhibit
convergence to Brownian motion, for circular and even randomly oriented square scatterers, for which
there is no exponential separation of initial conditions~\cite{DC00,DC01,LWH02}.  For the randomly
oriented squares in the open case (sufficiently large
fixed size and time increasing) it is clear that escape is $C/t$ (from period three orbits in acute triangles), so
there is likely some combination of limits $t\to\infty$, $L\sim t^\alpha$ for $0\leq\alpha\leq1/2$ at which
there is a transition from anomalous to normal diffusive behaviour.

\begin{problem}
Does the non-overlapping random Lorentz gas have convergence to Brownian motion?
\end{problem}

\section{Applications}
In this final section we summarize the impact that the study of the Lorentz gas has on other fields, past, present
and future, drawing together threads from the previous sections.

\paragraph{Probability} As discussed in Sec.~\ref{s:DIH},
the infinite horizon Lorentz gas, has been a prime example of non-standard
convergence to the normal distribution, that is, with logarithmic scaling in time.  It is the venue in which the
anomalous convergence of moments was discovered, and is currently under investigation.  Models with weaker
correlations, namely flat points, Sec.~\ref{s:flat} exhibit anomalous Burnett coefficients, which are relevant more
generally to rate of convergence in local limit theorems.
\paragraph{Dynamical systems} The Boltzmann-Sinai ergodic hypothesis~\cite{Szasz00} provided much of the original
impetus for ergodic theory, and has only recently been resolved in its original form,~Sec.~\ref{s:MD}.
Dispersing Lorentz gases, particularly in higher dimensions, provide a continuing challenge to the study of
hyperbolic dynamics with singularities~\cite{BT12}, while polygonal models have spurred and made accessible
the recently active field of flows on flat surfaces with singularities, Sec.~\ref{s:poly}.  Surfaces of infinite
genus, corresponding to billiards with irrational angles, remain a major challenge.
\paragraph{Statistical physics} The Lorentz gas has provided a useful model of transport, both diffusion and heat conduction,
in that a single moving particle exhibits many features of the full (multi-particle) problem.  It was possible to prove
validity of the relevant (linear) Boltzmann equation, Sec.~\ref{s:random}, as well as providing
a simpler context to investigate logarithmic terms in the low density expansion of the diffusion
coefficient~\cite{vLW67}.  More recently it has
elucidated many of the connections between microscopic dynamics (for example reversibility, Lyapunov exponents,
dimensions) and macroscopic transport (for example irreversibility, transport coefficients).  See Secs.~\ref{s:ext}
and~\ref{s:esc} and also Refs.~\cite{D00,DC00,Gaspard05}.
\paragraph{Molecular simulation} The periodic Lorentz gas is equivalent to two-particle molecular dynamics
with periodic boundary conditions, following centre of mass reduction.  It has been used as a test-bed for properties
of thermostats, additional terms in the equations of motion that take account of effects of the environment, Sec.~\ref{s:gauss}.
The periodic boundary conditions, while helpful for studying bulk effects, can lead to substantial
modifications of the dynamical properties, particularly at low densities, Sec.~\ref{s:MD}.
\paragraph{Physics of transport}  Finally, the Lorentz gas and similar models have often been used to model
transport on small scales.  In this context, the use of polygonal channels for studying nanopores was mentioned in
Sec.~\ref{s:poly}.  Lorentz channels have been used to understand thermoelectric efficiency~\cite{BC11}.  Other
examples have included confined fluids~\cite{KMS11,SSAHD13}, glasses~\cite{VH09}, nuclear collisions~\cite{BG11}
and zeolites~\cite{KC03}.

\section*{Acknowledgements} The author is grateful for the kind hospitality of the Kavli Institute for Theoretical
Physics, China, in July 2013, and also BME Budapest in January 2012, where the work presented in Sec.~\ref{s:flat}
was commenced.  He is also grateful for helpful information and discussions with Peter B\'alint (in particular for
Sec.~\ref{s:flat}), Giampaolo Cristadoro, Daniel El-Baz, Pierre Gaspard, Eivind Hauge, Rainer Klages, Paul Krapivsky,
Jens Marklof, Gary Morriss, David Sanders
(in particular for Sec.~\ref{s:MD}), Andreas Str\"ombergsson, Domokos Sz\'asz,  B\'alint T\'oth, Corinna Ulcigrai
and Henk van Beijeren.

\bibliographystyle{custom3}
\bibliography{billiards}
%\printbibliography
\end{document}